 \documentclass[preprint,onecolumn,3p,times,sort&compress,12pt]{elsarticle}

\usepackage{amssymb}
\usepackage{hyperref}
\usepackage{breakurl,epstopdf,xcolor}

\journal{Carbon}

	\usepackage{fancyheadings}
\renewcommand{\thispagestyle}[1]{} 

\begin{document}

\begin{frontmatter}



\title{Ferrimagnetic and antiferromagnetic phase in bilayer graphene nanoflake controlled with external electric fields}


\author{Karol Sza{\l}owski}

\address{Department of Solid State Physics, Faculty of Physics and Applied Informatics, University of {\L}\'{o}d\'{z}, \\ul. Pomorska 149/153, 90-236 {\L}\'{o}d\'{z}, Poland}

\date{\today}

 \cortext[cor1]{Corresponding author. E −mail address:
kszalowski@uni.lodz.pl}

\begin{abstract}
The paper presents a computational study of the ground-state magnetic phases of a selected bilayer graphene nanoflake in external electric field and magnetic field. The electric field has parallel and perpendicular component while the magnetic field is oriented in plane. The system consists of two rectangular layers having armchair edges and zigzag terminations with Bernal stacking. The theoretical model is based on a tight binding Hamiltonian with Hubbard term. The magnetic phase diagram involving the total spin is constructed, showing the stability areas of phases with total spin values equal to 0 and 1. A significant stability range of antiferromagnetic, layer-like arrangements is found and extensively discussed. The possibility of switching between nonmagnetic, antiferromagnetic and ferrimagnetic phases with both components of external electric field is demonstrated, being a manifestation of a magnetoelectric effect. The influence of magnetic field on the phase diagrams is analysed.  

\vspace*{0.3cm}
\includegraphics[width=0.9\textwidth]{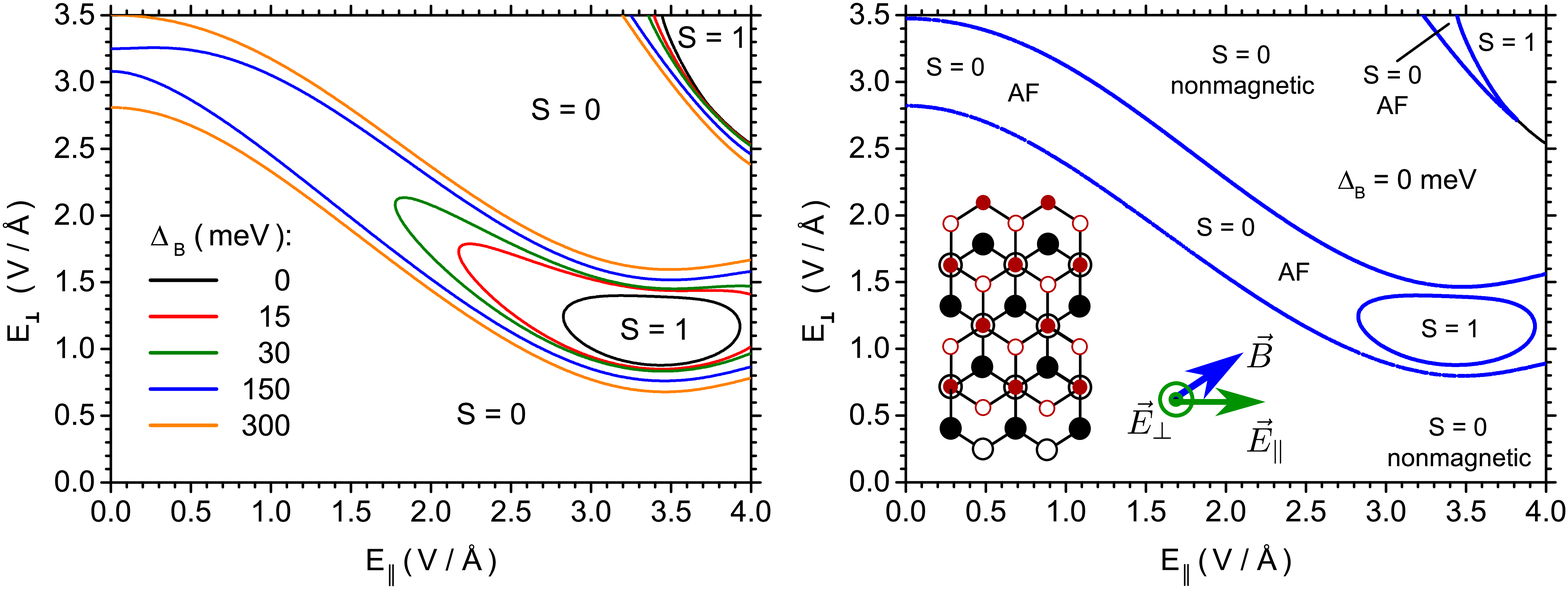}
\end{abstract}


\end{frontmatter}

\section{Introduction}

The discovery of graphene \cite{novoselov_electric_2004} intensified the interest in broad variety of carbon nanostructures \cite{georgakilas_broad_2015}. Among various low-dimensional carbon-based systems, bilayer graphene attracts considerable attention. The infinite bilayer graphene, unlike monolayer one, exhibits a gap in the electronic spectrum \cite{rozhkov_electronic_2016,mccann_electronic_2013}, making the connection with semiconducting materials used in classical electronics. What is more, it offers the possibility of breaking the inversion symmetry by applying a transverse electric field (perpendicular to the layers) in order to modify the electronic structure \cite{varlet_tunable_2015,sui_gate-tunable_2015,castro_electronic_2010,zhang_direct_2009,falkovsky_screening_2009,min_textitab_2007,
ohta_controlling_2006}. This possibility can be exploited with a view to tuning the gap and other properties to design devices. 

Within the broad field of graphene physics, its magnetism offers particularly exciting prospects for developing carbon-based spintronics \cite{han_graphene_2014}. As a result, also magnetic properties of bilayer graphene are widely explored in the literature   \cite{nafday_controlling_2016,klier_electrical_2016,jiang_study_2015,mohammadi_rkky_2015,nafday_magnetism_2013,lipinski_kondo_2013,castro_effect_2011,killi_controlling_2011}. In addition to the infinite, two-dimensional bilayers, bilayer graphene nanostructures constitute a class of highly promising systems due to additional possibility of shaping their edge geometry to engineer the desired properties. This concerns, for example, one-dimensional nanoribbons \cite{vu_modulation_2016,masrour_magnetic_2017,chung_electronic_2016,chang_configuration-dependent_2014,pan_noncollinear_2013,leon_half-metallicity_2013,li_edge_2012,sahu_energy_2008}. It is also notable that highly non-trivial magnetic properties of bilayer zero-dimensional nanoflakes, in particular influenced by electric field, attracted considerable attention \cite{guclu_sublattice_2016,krompiewski_edge_2016,rut_trigonal_2016,da_costa_magnetic_2016,farghadan_giant_2016,guclu_electric-field_2011, sahu_effects_2010}. Due to the presence of edge, magnetism can emerge in derivative graphene nanostructures even without magnetic impurities \cite{yazyev_emergence_2010}, while for infinite systems the adatoms or point defects are necessary for that purpose \cite{vejpravova_magnetic_2016}.   

In spite of well-motivated search for ferromagnetism in graphene, not only ferromagnetic ordering may exhibit potential for applications. Recently, spintronics based on antiferromagnetic materials  \cite{baltz_antiferromagnetism:_2016,jungwirth_antiferromagnetic_2016,gomonay_spintronics_2014} has emerged as a rapidly developing field owing to advantages guaranteed by using this class of magnetics, like insensitivity to external magnetic fields and lack of stray fields. The key feature necessary for applications in such spintronics is controlling the antiferromagnetic order by means of electric field. Therefore, among various magnets, those controllable with electric field appear most promising. Such an influence of the electric field on magnetic properties is a manifestation of magnetoelectric effect \cite{khomskii_coupled_2016,wang_multiferroicity:_2009,rivera_short_2009,eerenstein_multiferroic_2006,fiebig_revival_2005} which has attracted so far also some interest in the context of graphene systems \cite{santos_electric_2015,luo_bias_2015,zulicke_magnetoelectric_2014,santos_magnetoelectric_2013,zhang_tuning_2010,jung_magnetoelectric_2010,zhang_magnetoelectric_2009}. The search for novel platforms for spintronic devices drives the works aimed at exploring the physics of new magnetic systems. As a consequence, the interest in antiferromagnetic orderings in graphene-based systems and their control with electric field has a sound motivation. 

In our paper we present a computational study of magnetic orderings in a selected bilayer graphene nanostructure (nanoflake) in external in-plane and perpendicular electric field as well as in-plane magnetic field. Being interested in magnetoelectric effect, we explore the total spin phase diagram and identify a wide stability area of antiferromagnetic ordering within the zero total spin range. We investigate the possibility of switching between states with various total spin as well as between antiferromagnetic and non-antiferromagnetic orderings with electric fields and discuss the robustness of the effect versus the magnetic field. The theoretical model which we use to describe our system and the presentation of obtained results with their discussion will be shown in the next sections.

\section{Theoretical model}

\begin{figure*}[h!]
  \begin{center}
   \includegraphics[width=\textwidth]{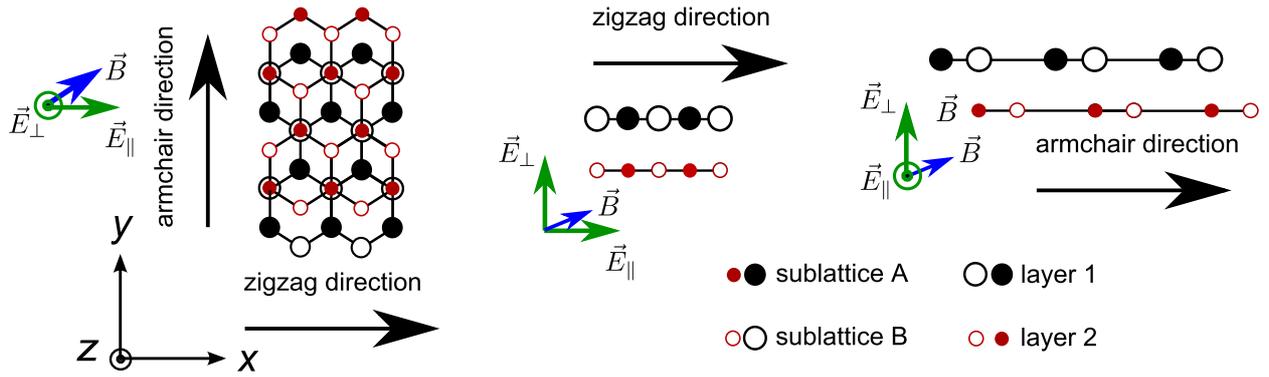}
  \end{center}
   \caption{\label{fig:1} Schematic view of the bilayer rectangular graphene nanoflake with the directions of applied electric and magnetic fields marked. Top view (left) as well as two side views (right) are presented. Carbon atoms belonging to different sublattices and two layers are marked with distinct symbols (color online).}
\end{figure*}

The system of interest in the present study is the bilayer graphene nanoflake with Bernal stacking \cite{rozhkov_electronic_2016} (for which both layers are shifted with respect to each other, so that only every second carbon atom in one layer is above the atom in the second layer). Both its layers are rectangular in shape, with armchair side edges and zigzag terminations, and can be treated as short sections of armchair graphene nanoribbon. We focus our attention on the particular structure composed of $N=40$ carbon atoms (with edges passivated for example by hydrogenation). The schematic top and side view of the system is presented in Fig.~\ref{fig:1}. As the graphene honeycomb lattice is bipartite, we denote the atoms belonging to both sublattices, A and B, with filled and empty circles, respectively. The charge neutrality state, i.e. one electron per carbon atom, is assumed. The nanoflake is immersed in external electric and magnetic field. The magnetic field $\vec{B}$ (which can be also the exchange field originating from the magnetic substrate) has arbitrary in-plane orientation. The electric field $\vec{E}$ is assumed to possess two components: in-plane one, $E_{\parallel}$, oriented along the zigzag terminations ($x$ axis, see Fig.~\ref{fig:1}) and perpendicular one, $E_{\perp}$, oriented perpendicularly to the nanoflake (along $z$ axis, see Fig.~\ref{fig:1}).

In order to characterize the magnetic behaviour of the system, we focus our interest on p$^{z}$ electrons. In order to model them, we utilize the following Hamiltonian:
\begin{eqnarray*}
\label{eq1}
\mathcal{H}&=&\,-\sum_{\left\langle i,j\right\rangle,\sigma}^{}{t_{ij}\,\left(c^{\dagger}_{i,\sigma}\,c^{}_{j,\sigma}+c^{\dagger}_{j,\sigma}\,c^{}_{i,\sigma}\right)}+U\sum_{i}^{}{ \left(\left\langle n_{i,\uparrow} \right\rangle n_{i,\downarrow} +\left\langle n_{i,\downarrow} \right\rangle n_{i,\uparrow}\right) }\\&&-U \sum_{i}^{}{\left\langle n_{i,\uparrow} \right\rangle \left\langle n_{i,\downarrow} \right\rangle}+\frac{1}{2}\Delta_{B}\displaystyle\sum_{i}^{}{s^{z}_{i}}+e\,E_{\parallel}\sum_{i,\sigma} ^{}{x_i n_{i,\sigma}}+e\,E_{\perp}\sum_{i,\sigma} ^{}{z_i n_{i,\sigma}}. 
\end{eqnarray*}

It is composed of tight-binding part, describing the electron hopping between nearest-neighbouring carbon atoms labelled with $i$ and $j$, both in-plane (with intraplanar hopping integral $t_{ij}=t_{\parallel}$ = 3.0 eV \cite{zhang_determination_2008}) and between the planes (with interplanar hopping integral $t_{ij}=t_{\perp}$= 0.4 eV \cite{zhang_determination_2008}). The operators $c^{\dagger}_{i,\sigma}$ ($c^{}_{i,\sigma}$) create (annihilate) an electron at site $i$ with spin $\sigma=\uparrow,\downarrow$. The tight-binding part is supplemented with Hubbard term with $U$ being an effective on-site Coulombic interaction energy between the electrons and the number of electrons is provided by the operator $n_{i,\sigma}=c^{\dagger}_{i,\sigma}\,c^{}_{i,\sigma}$. The Hubbard term is considered within Mean Field Approximation \cite{szalowski_graphene_2015}. Further terms are responsible for introducing the external fields. Namely, the effect of the in-plane magnetic field is described by means of Zeeman splitting energy $\Delta_{B}$ with $z$ component of spin at site $i$ expressed by $s_{i}^{z}=\left(n_{i,\uparrow}-n_{i,\downarrow}\right)/2$. In addition, the electric field is described by its potential acting on the total charge $e\sum_{\sigma}^{}{n_{i,\sigma}}$ at site $i$ (with $e$ being an elementary charge), what is reflected by two terms - separately for in-plane field $E_{\parallel}$ along $x$ direction and for perpendicular field $E_{\perp}$ along $z$ direction. 

The further treatment of the Hamiltonian is aimed at description of the ground state of the system in question (at the temperature $T=0$). The Hamiltonian is diagonalized self-consistently after splitting to the form of $\mathcal{H}=\mathcal{H}_{\uparrow}+\mathcal{H}_{\downarrow}-U\sum_{i}^{}{\left\langle n_{i,\uparrow} \right\rangle\left\langle n_{i,\downarrow} \right\rangle}$. The whole procedure follows the detailed description presented in our work Ref.~\cite{szalowski_graphene_2015}. As a result, the single-particle energy eigenvalues as well as eigenvectors are obtained, allowing to calculate the total energy and all other ground-state quantities (such as total spin $S$ or spin distribution over the nanostructure). In particular, all the solutions corresponding to the total number of electrons equal to $N$ (while the number of spin-up and spin-down electron varies) are tested. The solution with the lowest energy is accepted as a ground-state one. Moreover, the procedure is repeated starting from random initial conditions (charge distribution) in search for the lowest obtained total energy. 

In order to characterize the magnetic state of the system, the crucial quantity is the total spin $S=\displaystyle \sum_{i}^{}{s_{i}^{z}}$. However, for exploration of possible antiferromagnetic (AF) phases, staggered order parameters can be introduced. As a rule, the antiferromagnetic system can be subdivided into two subsystems within which the spins exhibit parallel orientation. The difference of magnetizations of both subsystems serves as an antiferromagnetic order parameter. For the system of interest, two ways of making such division can be imagined. One of them is identification of each subsystem with one layer of the flake. As a consequence, a staggered magnetization is the difference of layer magnetizations: $\displaystyle m_{layer}=\left|\sum_{i\in \mathrm{ layer \,\,1}}^{}{s_{i}^{z}}-\sum_{i\in \mathrm{ layer\,\, 2}}^{}{s_{i}^{z}}\right|$. Another division can be based on identification of each subsystem with a single sublattice, A or B. Therefore, a staggered magnetization is then a difference is sublattice magnetizations, $\displaystyle m_{sublattice}=\left|\sum_{i\in \mathrm{ sublattice\,\,A}}^{}{s_{i}^{z}}-\sum_{i\in \mathrm{ sublattice\,\, B}}^{}{s_{i}^{z}}\right|$.

The calculated values of ground-state total spin are used further to construct the magnetic phase diagrams for the bilayer nanoflake. Moreover, the knowledge of spin distribution over particular sites of the nanostructure enables the search for the antiferromagnetic arrangements and detailed analysis of the behaviour of antiferromagnetic order parameters as a function of the external fields. Such numerical results will be presented in the following section. The effect of electric field will be particularly emphasized.

\section{Numerical results}

\begin{figure}[h!]
  \begin{center}
   \includegraphics[width=0.7\columnwidth]{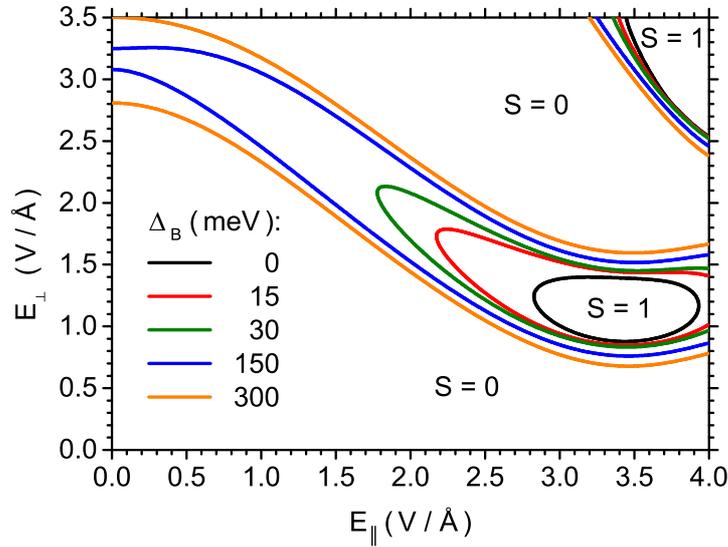}
  \end{center}
   \caption{\label{fig:2} Ground-state total spin phase diagram for the considered bilayer nanoflake. The borders between stability areas of phases with total spin $S=0$ and $S=1$ are plotted with solid lines as a function of in-plane and perpendicular electric field, for various splitting energies (color online).}
\end{figure}

\begin{figure*}[h!]
  \begin{center}
   \includegraphics[width=\textwidth]{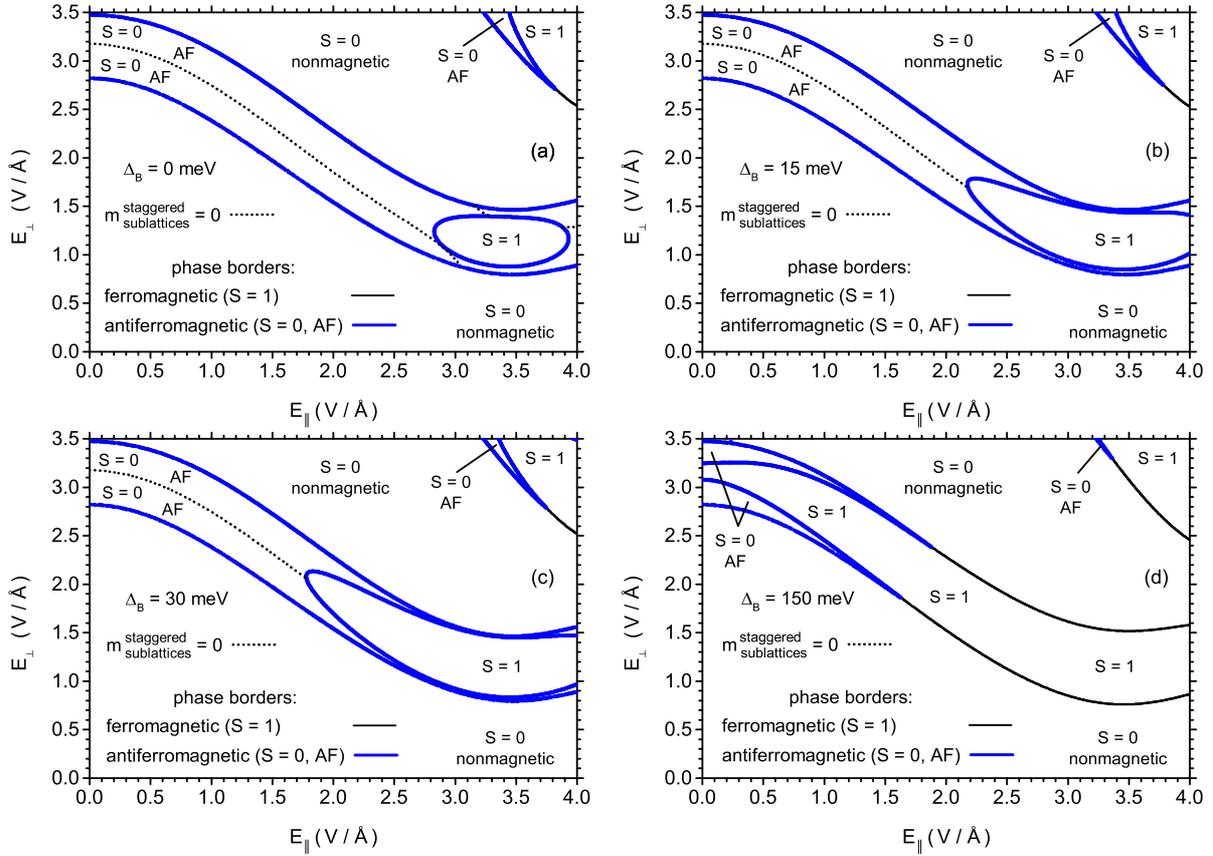}
  \end{center}
   \caption{\label{fig:3} Ground-state phase diagram for the considered bilayer nanoflake, emphasizing the stability areas of antiferromagnetic ordering, for various, increasing splitting energies. The borders between AF phase and other phases are plotted with thick solid lines. The AF order parameter for sublattices vanishes additionally along thin dotted lines.}
\end{figure*}

\begin{figure}[h!]
  \begin{center}
   \includegraphics[width=0.7\columnwidth]{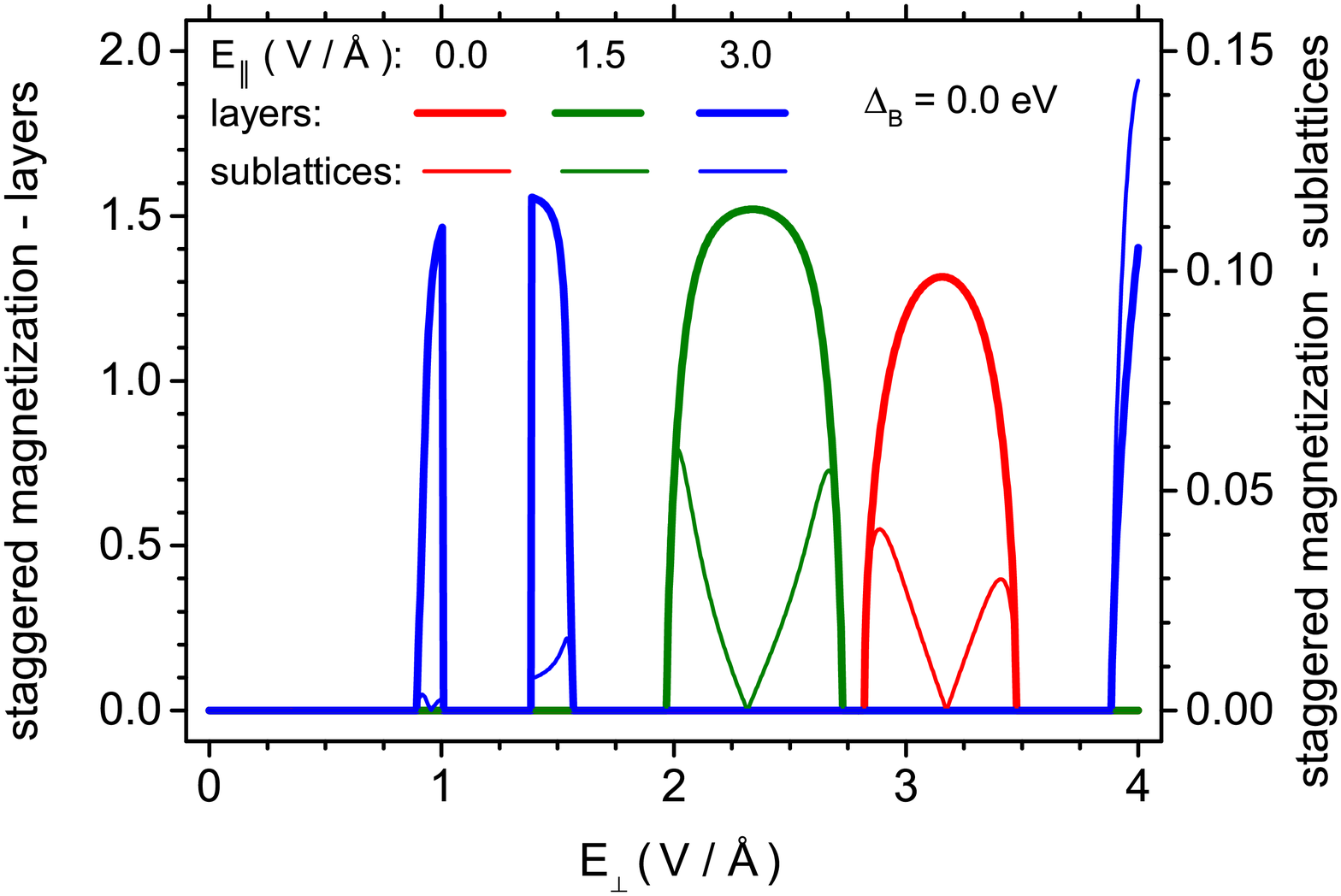}
  \end{center}
   \caption{\label{fig:4} The dependence of staggered magnetization for layers (left scale) and sublattices (right scale) on the perpendicular electric field, for three selected values of constant in-plane electric field and in the absence of the magnetic field (color online).}
\end{figure}

\begin{figure}[h!]
  \begin{center}
   \includegraphics[width=0.7\columnwidth]{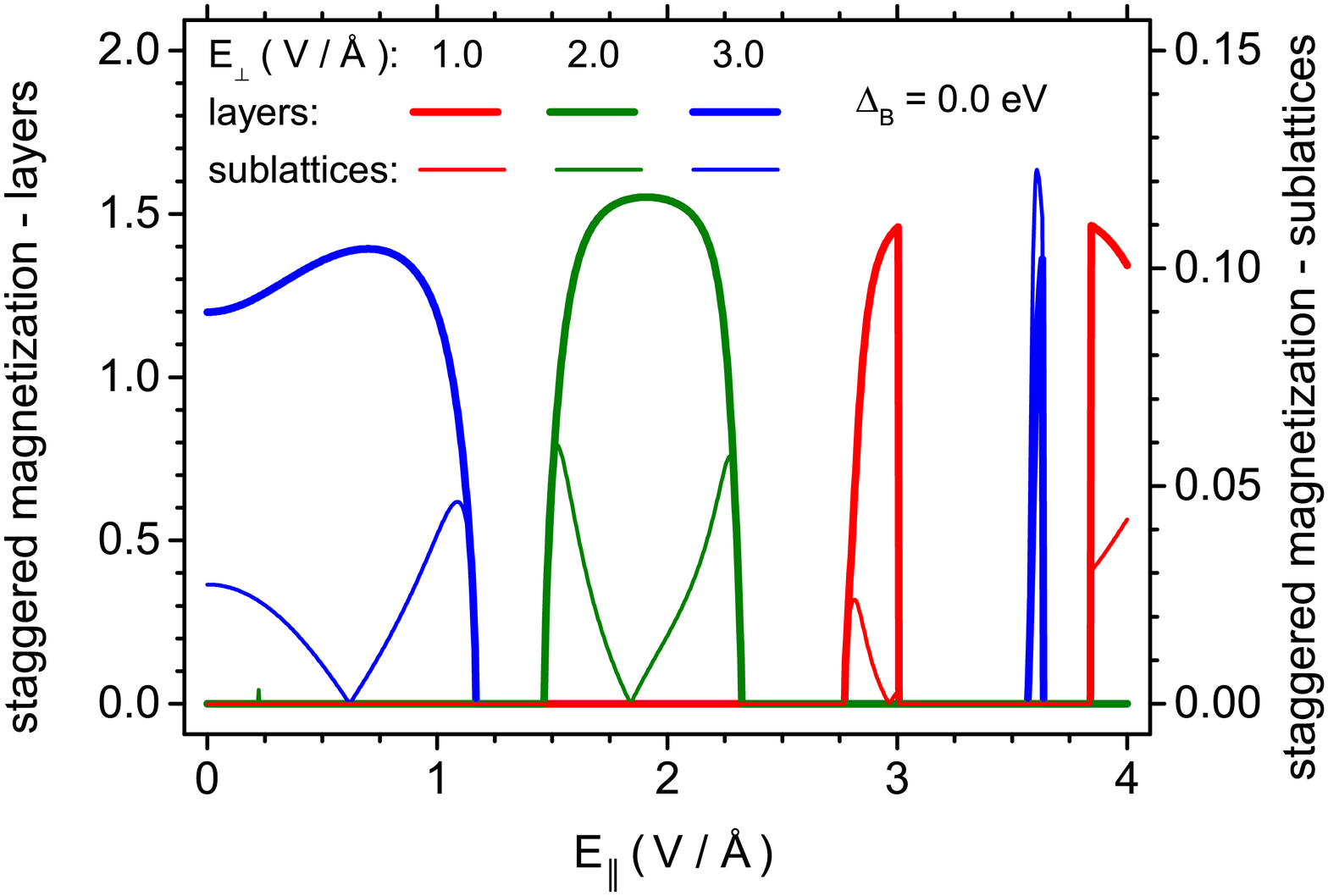}
  \end{center}
   \caption{\label{fig:5} The dependence of staggered magnetization for layers (left scale) and sublattices (right scale) on the in-plane electric field, for three selected values of constant perpendicular electric field and in the absence of the magnetic field (color online).}
\end{figure}

\begin{figure}[h!]
  \begin{center}
   \includegraphics[width=0.7\columnwidth]{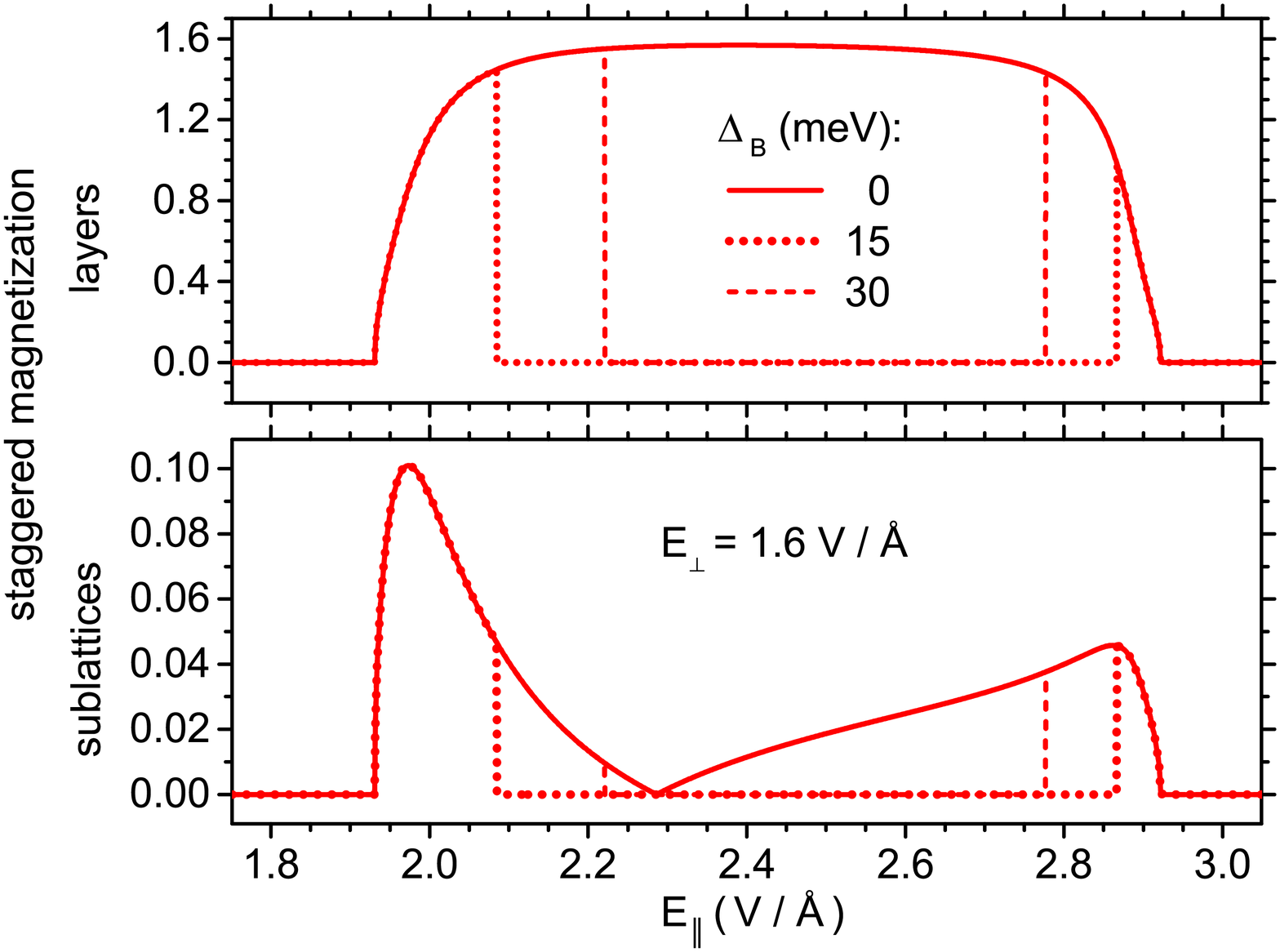}
  \end{center}
   \caption{\label{fig:6} The dependence of staggered magnetization for layers and sublattices on the in-plane electric field, for constant perpendicular electric field and for three selected values of splitting energy. }
\end{figure}

\begin{figure}[h!]
  \begin{center}
   \includegraphics[width=0.7\columnwidth]{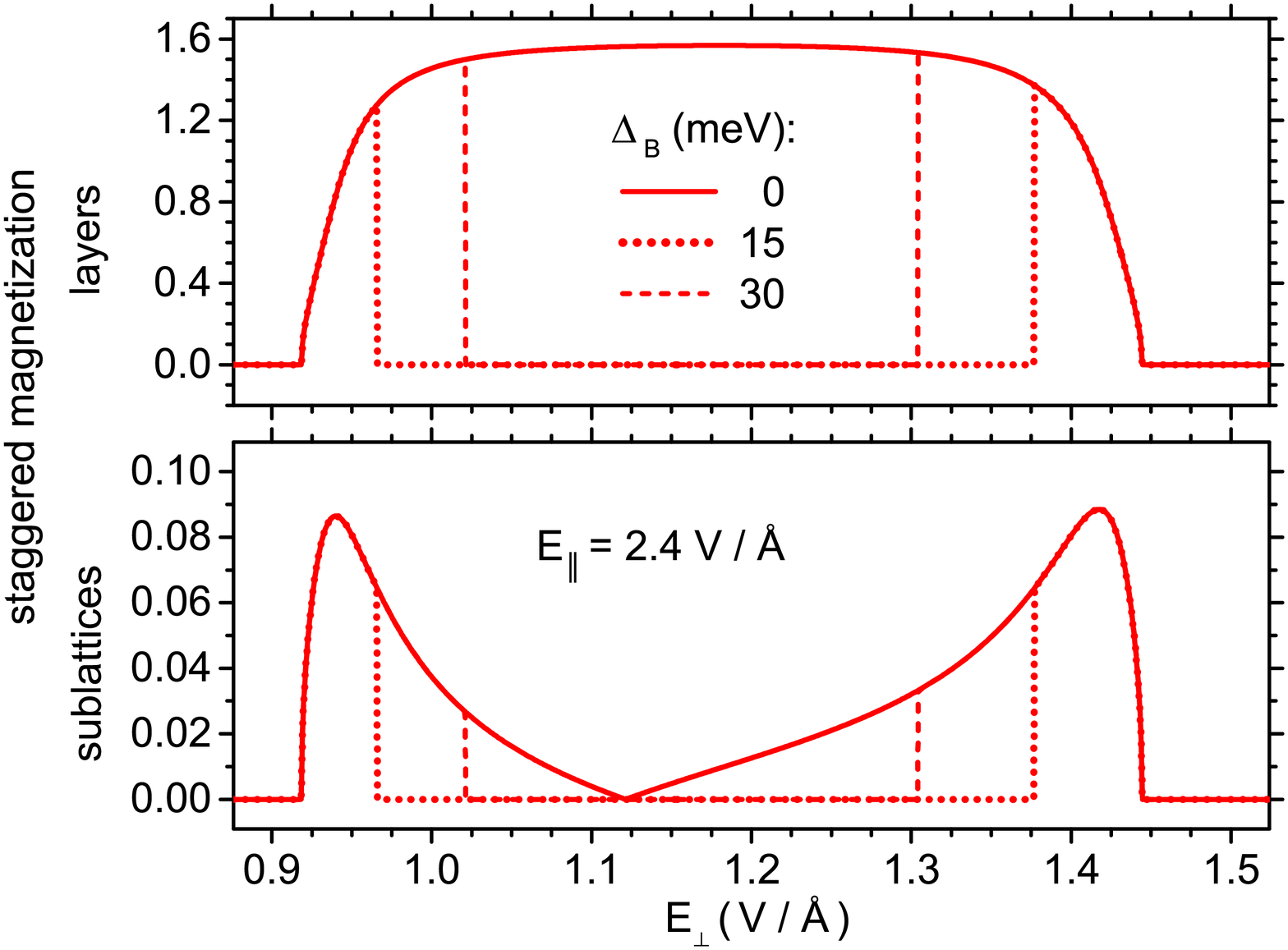}
  \end{center}
   \caption{\label{fig:7} The dependence of staggered magnetization for layers and sublattices on the perpendicular electric field, for constant in-plane electric field and for three selected values of splitting energy.}
\end{figure}

\begin{figure}[h!]
  \begin{center}
   \includegraphics[width=0.385\columnwidth]{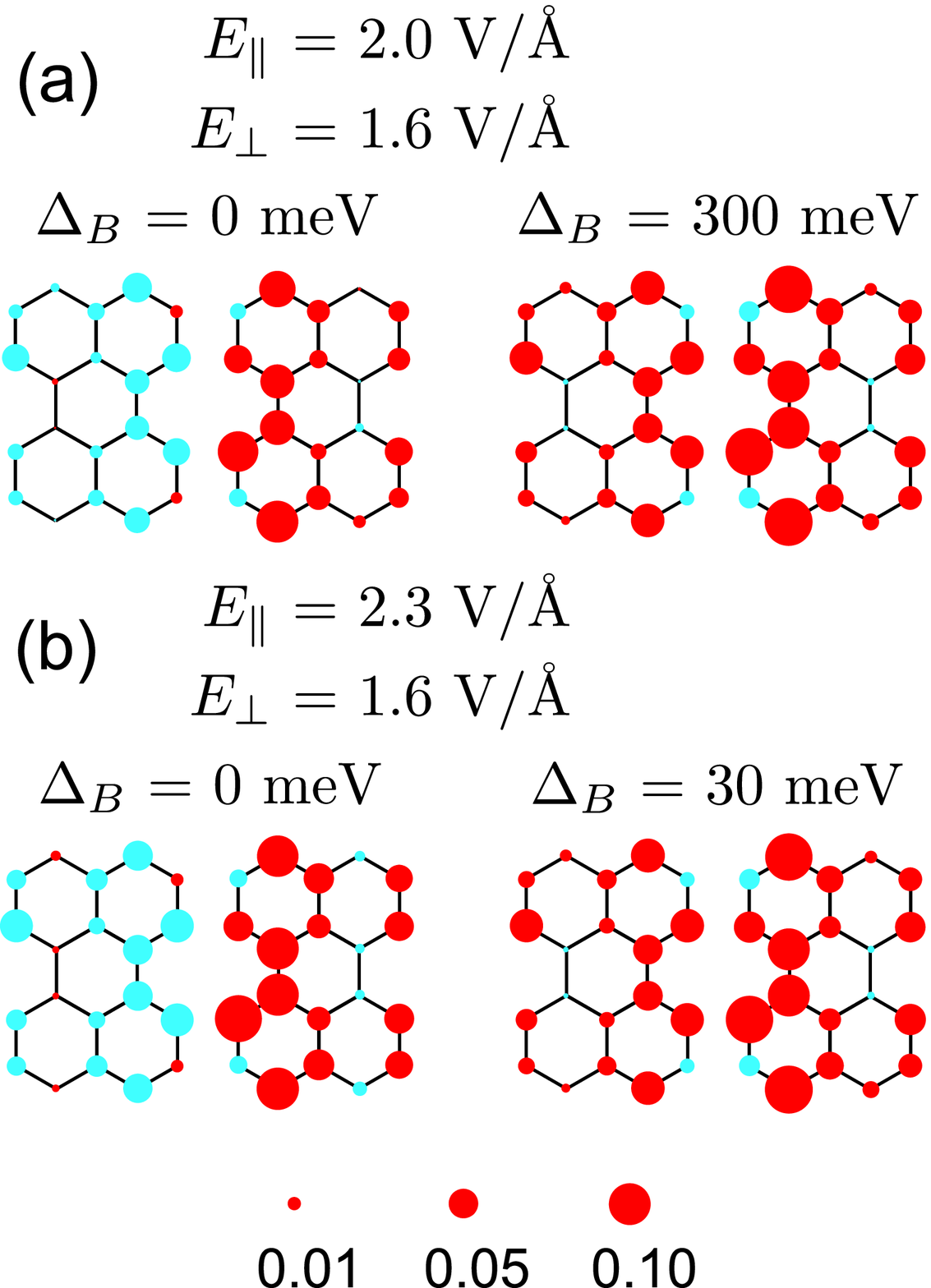}
  \end{center}
   \caption{\label{fig:8} The distribution of spin density across both layers of the nanoflake, for given values of both components of electric field and splitting energy. Two colors (color online) of the circles represent opposite spin orientations. The area of each circle is proportional to the spin density at given site. Left half of each panel represents the upper layer 1 of the bilayer nanoflake, while the right one is the lower layer 2. The circles corresponding to three characteristic given values of spin density are showed below the plot.}
\end{figure}

In order to characterize the sensitivity of the magnetic properties of the considered nanoflake to electric field (being a manifestation of the magnetoelectric effect), it is instructive to explore first the total ground-state spin as a function of in-plane and perpendicular component of the electric field. Such a ground-state magnetic phase diagram is presented in Fig.~\ref{fig:2}, showing the boundaries between the areas with the total $z$ component of the nanoflake spin equal to 0 and equal to 1 as a function of parallel and perpendicular component of the electric field. Let us comment here that the state with total spin equal to 1 is not a saturated ferromagnetic state, therefore, it will be called further a ferrimagnetic state. The plot shows the calculations performed for various values of Zeeman energy $\Delta_{B}$, ranging from 0 (no magnetic field) to 300 meV. It is evident that, in the absence of external magnetic field, the vast majority of the phase diagram corresponds to total spin equal to 0. Only an island of phase with spin 1 emerges for significant parallel electric field and moderate perpendicular field. Another such range is present at high values of both fields. Switching on Zeeman splitting results in expansion of the range of ferrimagnetic arrangement with $S=1$. For the high-field area in the upper right corner of the diagram, the changes due to Zeeman splitting are negligible. On the contrary, the second range of ferrimagnetic order is much more sensitive to $\Delta_{B}$ and expands along the diagonal of the diagram. It is noticeable that for the Zeeman splitting values of 150 meV the range with $S=1$ can be reached even at $E_{\parallel}=0$ by application of sufficiently strong $E_{\perp}$. 

The analysis of the total-spin phase diagram (Fig.~\ref{fig:2}) proves the presence of the magnetoelectric effect and shows the possibility of controlling the total spin of the nanostructure with external electric field, for both in-plane and perpendicular orientation. As a consequence, all-electric switching of the magnetic state could be achieved. In spite of the presence of stability range for $S=1$ phase, still a significant area of the diagram corresponds to total spin equal to 0. In such range the antiferromagnetic arrangements may potentially be stable. 

The following part of the section will be mainly focused on the properties of antiferromagnetic state. At the beginning, it is crucial to identify the stability range of antiferromagnetic ordering. 
The phase diagram analogous to Fig.~\ref{fig:2}, but showing the stability range of AF phase, is shown in Fig.~\ref{fig:3}(a)-(d). The part Fig.~\ref{fig:3}(a) corresponds to the absence of Zeeman splitting. It should be emphasized that the diagram was constructed on the basis of the staggered magnetization for layers $m_{layer}$. A noticeable stability range of antiferromagnetic phase was found. It is visible that the boundaries between AF phase and non-magnetic phase bear some resemblance to the boundaries between $S=0$ and $S=1$ ordering at high Zeeman splitting (compare with Fig.~\ref{fig:2}). In particular, the AF ordering range encircles the island of $S=1$ ordering for strong $E_{\parallel}$ and moderate $E_{\perp}$. In the vicinity of high-field boundary between $S=0$ and $S=1$, one more, wedge-shaped AF range is found. It is also interesting that another staggered magnetization parameter, $m_{sublattice}$, indicates the analogous behaviour as $m_{layer}$, taking the value of zero simultaneously with $m_{layer}$. However, $m_{sublattice}$ vanishes also in the middle of AF range, what was indicated additionally with dotted lines visible in Fig.~\ref{fig:3}(a-c).

It can be concluded that AF ordering is quite robust in the studied system and extends along the diagonal of the diagram. The transitions between AF phase and other phases can be realized by varying both components of the electric field, $E_{\parallel}$ and $E_{\perp}$.

In the presence of increasing Zeeman splitting, the ferrimagnetic phase can be expected to expand at the cost of AF stability range. This process can be followed in Fig.~\ref{fig:3}(b-d), showing the phase diagram with AF orderings indicated for increasing values of $\Delta_{B}$, equal to 15, 30 and 150 meV, respectively. The island of $S=1$ expands significantly along the diagonal of the diagram (see Fig.~\ref{fig:2}), pushing away the AF phase. This variability concerns only the shift of the boundary between AF and $S=1$ phase. However, it must be stated that the external boundary of AF stability range, corresponding to the transition between AF and non-magnetic phase is rather insensitive to $\Delta_B$. This can be especially confirmed by visual inspection of Fig.~\ref{fig:3}(d), presenting the phase diagram with AF orderings for strong magnetic field, $\Delta_B$= 150 meV. For such strong magnetic field, the stability range of AF phase splits to two separate ranges, present only for low parallel field and quite high perpendicular field. However, the external boundaries remain untouched. A limited, wedge-like, range of AF ordering for strong parallel and perpendicular electric field, in touch with another $S=1$ range, also tends to reduce its area when $\Delta_{B}$ increases.

In order to analyse the detailed behaviour of the antiferromagnetically ordered phases in the system of interest, it is worthy to study the cross-sections of the phase diagrams from Fig.~\ref{fig:3}. An example of such a cross-section is Fig.~\ref{fig:4}, which shows the dependence of both AF order parameters - $m_{layer}$ and $m_{sublattice}$ - on perpendicular electric field for three constant values of parallel field (i.e. the vertical cross-section of the phase diagram in Fig.~\ref{fig:3}(a)). The first feature, which can be observed for two lowest values of $E_{\parallel}$, is that both order parameters vary continuously with $E_{\perp}$, speaking especially about the behaviour at the border between AF phase and magnetically unordered phase. This proves that the transitions between the mentioned phases - AF one and completely magnetically unordered one -  are of second-order. Both $m_{layer}$ and $m_{sublattice}$ acquire non-zero values at the same points. As it was mentioned already, $m_{sublattice}$ vanishes also at some point where $m_{layer}>0$ (see dotted lines at Fig.~\ref{fig:3}(a-c)). The cross-section plotted for the highest considered value of $E_{\parallel}$ covers also the borders between the AF phase and ferrimagnetic phase with spin $S=1$. It is visible that at these points both AF order parameters vary discontinuously (in the same manner as the total spin, wich is switched between 0 and 1). Therefore, the border between AF state and ferimagnetic state corresponds to the first-order phase transition. 

Another interesting feature is the significant difference in magnitude of both AF order parameters, as generally $m_{sublattice}$ is about an order of magnitude smaller than $m_{layer}$ (please note double vertical scale in Fig.~\ref{fig:4}). The order parameter for layers achieves the values close to one or more, therefore, the antiferromagnetic ordering is rather layer imbalance-based. 

A similar cross-sections, but performed for constant $E_{\perp}$, are shown in Fig.~\ref{fig:5} (presenting horizontal cross-sections of Fig.~\ref{fig:3}(a)). In particular, the plot for lowest considered perpendicular field exhibits two ranges of AF ordering separated by ferrimagnetic phase ($S=1$). It confirms the discontinuous character of transition between such orderings. Moreover, also vanishing of sublattice order parameter $m_{sublattice}$ is evident at some points with non-zero $m_{layer}$ value. The same difference in magnitude between $m_{layer}$ and $m_{sublattice}$ as described before can be noticed.

Analysis of the cross-sections of the phase diagram presented in Figs.~\ref{fig:4} and \ref{fig:5} brings the conclusion that the AF ordering can be rather easily controlled by both components of the electric field. Although the transitions between magnetically unordered phase and AF-ordered one are continuous in character, yet the variability of the staggered magnetizations $m_{layer}$ and $m_{sublattice}$ is rather rapid close to the phase boundary and noticeable values of order parameter are achieved quite fast. Let us remind that at the boundary between AF phase and $S=1$ phase the changes of order parameters are discontinuous in character. The antiferromagnetic orderings in the systems of interest are dominantly based on division of the system into layers.

From the point of view of the possible applications, it is particularly important to characterize the sensitivity of the AF orderings to the magnetic field. As a consequence, it is instructive to analyse selected cross-sections of the phase diagram presented in Fig.~\ref{fig:3}(b)-(d). First such cross-section is shown in Fig.~\ref{fig:6}, for constant field $E_{\perp}$ and variable field $E_{\parallel}$, for three values of the Zeeman splitting $\Delta_{B}$. Both layer and sublattice staggered magnetization are plotted. It can be clearly seen how the expansion of the 'island' of ferrimagnetic phase with $S=1$ (see Fig.~\ref{fig:3} pushes away the AF phase range, dividing it (in the cross-section) into two areas separated with $S=1$ ordering. However, when $\Delta_{B}$ increases, the values of both considered staggered magnetizations $m_{layer}$ and $m_{sublattice}$ are not altered within the range of antiferromagnetic phase. A fully similar behaviour is demonstrated in the second cross-section presented in Fig.~\ref{fig:7} for the case of constant field $E_{\parallel}$ and variable field $E_{\perp}$. In general, it can be concluded that the antiferromagnetic order is quite robust against the in-plane magnetic field.

Finally, it is interesting to discuss the distribution of spin density across the layers of the nanoflake for both antiferromagnetic and ferrimagnetic orderings. Such a distribution is shown in Fig.~\ref{fig:8} and corresponds to external fields  related to Fig.~\ref{fig:6} (i.e. $E_{\perp}$ value is the same as in Fig.~\ref{fig:6}). The case Fig.~\ref{fig:8}(a) shows the distribution for $E_{\parallel}=2.0$ V/\AA , what corresponds roughly to the maximum value of sublattice AF order parameter, while the case Fig.~\ref{fig:8}(b) is plotted for $E_{\parallel}=2.3$ V/\AA, where the order parameter $m_{sublattice}$ approximately vanishes. The left half of each plot is for the spin distribution in the upper layer 1, while the right one is for the lower layer 2. Moreover, the plot shows the results in the absence of magnetic field ($\Delta_B$=0) and in the presence of it ($\Delta_B>0$). The area of the circles is proportional to the spin density $s_{i}^{z}$ at the given site. It is clearly visible that all the distributions are rather asymmetric, what can be attributed to the presence of both in-plane and perpendicular electric field. In general, the spin densities in layer 2 are higher than in layer 1, what reflects the influence of perpendicular field. In the AF phase, a clearly dominant layer character of antiferromagnetism can be noticed, since almost all the spins in the same layer point in the same direction. The imbalance of spin densities for both sublattices is mush less pronounced. Let us remind here that $m_{layer}\gg m_{sublattice}$ as shown in Figs.~\ref{fig:4} and ~\ref{fig:5}.        
    
Let us observe that in the ferrimagnetic state (for $\Delta_{B}>0$) not all the sites indicate the same spin orientation (therefore it is not a saturated ferromagnetic state). It should be, however, stated that the total spin of the nanoflake can take values higher than one when increasing the magnetic field (i.e. $\Delta_{B}$) - see for example Fig.~1 in Ref.~\cite{szalowski_ground-state_2013}.

\section{Discussion}
\label{Discussion}
It might be mentioned that the antiferromagnetic orderings controlled with external field and other means in monolayer graphene nanostructures (nanoflakes, quantum dots, nanoribbons) were studied in Refs.~\cite{basak_optical_2016,zhu_magnetic_2016,szalowski_graphene_2015,li_electronic_2014,kabir_manipulation_2014,zhou_electric_2013,wang_topological_2009}. However, extending the system to bilayer one provides an additional possibility of applying independently in-plane and perpendicular electric field, enriching the phase diagram and enabling the extra control parameter. Moreover, gating of the system, leading to the presence of perpendicular electric field may be easier achievable from the experimental point of view.

The chance of experimental verification of the predictions is crucially dependent on the possibility of producing well-defined graphene nanoflakes with controllable edge geometry. In that context, let us mention both the work Ref.~\cite{konishi_investigating_2013} and especially Ref.~\cite{ruffieux_-surface_2016,wang_giant_2016} (all of them dealing with monolayer structures, but with rectangular shape and perfectly defined geometry). It is worth particular emphasis that the described properties are vitally susceptible to the form of wavefunctions in small, zero-dimensional systems with dominant edge contribution. The states corresponding to those predicted by means of a tight-binding model were detected experimentally and characterized with scanning tunneling microscopy/spectroscopy methods in geometrically well-defined structures Ref.~\cite{wang_giant_2016}. Also the spin polarization captured with tight-binding models with Hubbard term was proved to exist. The experimental progress in obtaining such rectangular nanoflakes served as one of motivations for selection of geometry of the system in the present study. It should be also mentioned that systems of larger in-plane extension could, on the one hand, exhibit richer phase diagram, but, on the other hand, the effect of screening of the in-plane electric field would become highly significant. Therefore, the smallest systems appear much more interesting from the point of view of studying the electric field-controlled magnetism. However, the increasing in-plane size is not an obstacle when the perpendicular field is used.

The in-plane magnetic field which we consider in our model should be rather regarded as an exchange magnetic field originating from the proximity of the graphene bilayer nanostructure to some magnetic substrate. The spin-dependent splittings of the energy bands are predicted and observed for numerous systems consisting of graphene on magnetic insulators \cite{leutenantsmeyer_proximity_2017,wei_strong_2016,hallal_tailoring_2016,swartz_integration_2012} and can reach, according to the calculations, the values of approximately one tenth of the hopping integral between the nearest neighbours in-plane. Even such large splittings were not predicted to fully destroy the AF ordering in our calculations. 

Last, but not least, it should be mentioned that the presence of antiparallel magnetic ordering at the zigzag edges of graphene nanoribbons was deduced from the experimental data and successfully modelled with Hubbard Hamiltonian also in the Ref.~\cite{magda_room-temperature_2014}. Regarding the applicability of the Mean Field Approximation to Hubbard model in the case of modelling the graphene nanostructures, its usefulness was discussed in Ref.~\cite{feldner_magnetism_2010}.

\section{Final remarks}
\label{Final remarks}
The rectangular bilayer nanoflake with Bernal stacking was shown to exhibit an interesting electric and magnetic field-dependent magnetic phase diagram in the ground state. Therefore, the presence of a clear magnetoelectric effect was demonstrated for the system in question. In the absence of magnetic field, the limited, ferrimagnetic range of total spin equal to 1 was found for noticeable values of both in-plane and perpendicular electric field. This range extends along the diagonal of the phase diagram under the influence of increasing in-plane magnetic field. However, within the range of total spin equal to 0, significant stability areas of antiferromagnetic phase were revealed. For the AF orderings mainly the spins of both layers of the flake are antiparallel, the imbalance of total spin between the sublattices is much weaker. The transitions between nonmagnetic states and states with AF ordering are continuous in character, while the transitions between AF states and ferrimagnetic state with total spin equal to 1 are discontinuous. The wide possibility of switching between the states with various magnetic properties was demonstrated. In particular, both in-plane and perpendicular components of electric field can be used for that purpose. The increase of external magnetic field reduces the stability area of AF ordering, however, within the AF phase the staggered magnetization values remain untouched.

The possible extensions of the presented study may involve systems of larger size and different shape or edge geometry (for example hexagonal ones, as in Ref.~\cite{new}). Also the investigation of finite temperature effects or charge doping would be valuable. We believe that our work makes a solid ground for further studies.

\section*{Acknowledgments}

\noindent The computational support on Hugo cluster at Laboratory of Theoretical Aspects of Quantum Magnetism and Statistical Physics, P. J. \v{S}af\'{a}rik University in Ko\v{s}ice is gratefully acknowledged.

\noindent This work has been supported by Polish Ministry of Science and Higher Education on a special purpose grant to fund the research and development activities and tasks associated with them, serving the development of young
scientists and doctoral students.

\end{document}